# A new spatial-scan thermoreflectance method to measure a broad range of anisotropic in-plane thermal conductivity

Puqing Jiang,[1,2,3,a),#] Dihui Wang,[2,3,#] Zeyu Xiang,[1,#] Ronggui Yang,[1] and Heng Ban[2,3]

[1]*School of Power and Energy Engineering, Huazhong University of Science and Technology, Wuhan, Hubei 430074, P.R. China*
[2]*Department of Mechanical Engineering and Materials Science, University of Pittsburgh, Pittsburgh, Pennsylvania 15261, USA*
[3]*Pittsburgh Quantum Institute, Pittsburgh, Pennsylvania 15260, USA*

**Abstract**

In-plane thermal conductivities of small-scale samples are hard to measure, especially for the lowly conductive ones and those lacking in-plane symmetry (i.e., transversely anisotropic materials). State-of-the-art pump-probe techniques including both the time-domain and the frequency-domain thermoreflectance (TDTR and FDTR) are advantageous in measuring the thermal conductivity of small-scale samples, and various advanced TDTR and FDTR techniques have been developed to measure transversely anisotropic materials. However, the measurable in-plane thermal conductivity ($k_{in}$) is usually limited to be $> 10 \text{ W/(m} \cdot \text{K)}$. In this work, a new spatial-scan thermoreflectance (SSTR) method has been developed to measure a broad range of $k_{in}$ of millimeter-scale small samples, including those lacking in-plane symmetry, extending the current limit of the measurable $k_{in}$ to as low as 1 W/(m·K). This SSTR method establishes a new scheme of measurements using the optimized laser spot size and modulation frequency and a new scheme of data processing, enabling measurements of in-plane thermal conductivity tensors of a broad range of $k_{in}$ values with both high accuracy and ease of operation. Some details such as the requirement on the sample geometry, the effect of the transducer layer, and the effect of heat loss are also discussed. As a verification, the $k_{in}$ of some transversely isotropic reference samples with a wide range of $k_{in}$ values including fused silica, sapphire, silicon, and highly ordered pyrolytic graphite (HOPG) have been measured using this new SSTR method. The measured $k_{in}$ agree perfectly well with the literature values with a typical uncertainty of 5%. As a demonstration of the unique capability of this method, the in-plane thermal conductivity tensor of x-cut quartz, an in-plane anisotropic material, has also been measured.

---

[#]These authors contributed equally.
[a]Corresponding authors. E-mail address: jpq2021@hust.edu.cn (P. Jiang)

## 1. Introduction

In-plane thermal conductivities of bulk and thin-film materials are of great interest in a wide variety of applications such as thermoelectrics[1], power electronics[2], and lithium-ion batteries[3]. Accurate measurements of the in-plane thermal conductivities of these materials are not only of great scientific value but also critically important for their engineering applications. In recent years, pump-probe techniques including both the time-domain thermoreflectance (TDTR) and frequency-domain thermoreflectance (FDTR) have become popular in measuring the anisotropic thermal conductivity of materials.[4-7]

TDTR is a powerful technique based on an ultrafast laser for thermal characterization.[8-10] The high modulation frequency of TDTR (in the MHz range) gives it many advantages such as short thermal diffusion length (and thus high spatial resolution) and high signal-to-noise ratio (SNR). However, because of the short thermal diffusion lengths (0.1~10 μm) compared to the laser spot size, TDTR is most readily sensitive to thermal conductivities in the cross-plane rather than the in-plane direction. Although several TDTR-based advanced approaches such as beam-offset TDTR[11, 12], variable spot size TDTR[13], and elliptical-beam TDTR[14, 15] have been developed to measure in-plane thermal conductivities ($k_{in}$), the measurable $k_{in}$ is usually limited to be >10 W/(m·K). Measuring lower $k_{in}$ would require a very low modulation frequency, which, however, is prohibited in TDTR due to the strong pulse accumulation[16].

FDTR is a variation of TDTR. Unlike TDTR, FDTR usually monitors the thermoreflectance signals as a function of the heating frequency.[17, 18] Most FDTR systems are based on continuous-wave lasers, which place no restrictions on the modulation frequency.[19] Theoretically, a very low modulation frequency can be used in FDTR to measure low in-plane thermal conductivities. In the literature, a beam-offset FDTR approach has been developed[20-23] to enhance the sensitivity of the frequency-dependent phase signals to the in-plane heat diffusion by offsetting the pump and probe spots at a fixed distance. Recently, Tang and Dames[24] extended the capability of beam-offset FDTR to measure the full thermal conductivity tensor of transversely anisotropic materials. However, one major difficulty of the FDTR experiments is that the reference phase of the lock-in amplifier needs to be carefully characterized to prevent the miscellaneous non-thermal phase shifts from interfering with the thermal phase signal. Besides, the current beam-offset FDTR approaches typically have their modulation frequencies limited to be >5 kHz, making this method only capable of measuring $k_{in} > 10$ W/(m·K).[22, 23] Although a transducerless FDTR approach has been developed to measure a lower $k_{in}$,[25] this approach could only apply to semiconductors with strong optical absorption and thus has a limited application.

Alternatively, a spatial-scan thermoreflectance method can also measure $k_{in}$. This method uses the same instrument as FDTR, and the phase signal as a function of pump-probe offset distance at a fixed modulation frequency is analyzed to derive $k_{in}$. Great progress has been achieved over the past twenty years in developing this spatial-scan thermoreflectance technique, alternatively also known as spatial-domain thermoreflectance[26], thermal wave microscopy[27], or modulated thermoreflectance (MTR)[28-31]. However, most of the prior efforts limit their applications to isotropic samples, and the potential of this technique to measure the full $k_{in}$ tensor of transversely anisotropic materials is not revealed. Besides, in the past efforts thus far, the absolute phase signals are analyzed to derive the thermal properties, which is still plagued by the same problem in FDTR that miscellaneous non-thermal phase signals easily interfere the desired thermal phase signal.

In this work, we further develop the spatial-scan thermoreflectance method with a unique capability of measuring the full $k_{in}$ tensor of small-scale transversely anisotropic samples over a broad range of $k_{in}$ from 1 to 2000 W/(m · K). Compared to the previous pump-probe techniques, this new SSTR method does not require characterizing the miscellaneous non-thermal phase shifts and thus is easier to implement and also less prone to error in the acquired signals. Following guidelines in choosing the appropriate laser spot size and modulation frequency for the measurements, signals acquired by this method are particularly sensitive to the in-plane thermal diffusivity $k_{in}/C$ of the substrate along the scanning direction but not sensitive to other parameters, thus yielding a small uncertainty in measuring the directional $k_{in}$ of the substrate.

This paper is organized as follows. First, details of the experimental setup and the basic measurement procedures are presented. Then more details regarding this method including the sensitivity analysis, the uncertainty estimation, and the effects of the transducer layer, sample geometry, and heat loss will be fully discussed. This method will then be verified by measuring in-plane thermal conductivities of some transversely isotropic reference samples including silica, sapphire, silicon, and HOPG. Finally, the capability of this method in measuring the full in-plane thermal conductivity tensor of transversely anisotropic materials will be demonstrated through both a simulated numerical experiment and a real measurement of x-cut quartz.

## 2. Methodologies

*2.1. Experimental setup and basic measurement procedures*

The improved spatial-scan thermoreflectance experiments have been conducted independently by different systems in two labs. The experimental system in the lab of the University of Pittsburgh is shown in Fig. 1(a). This system uses a 532 nm green laser as the pump beam and a 671 nm red laser as the probe beam, at which wavelength some low-thermal conductivity metals such as Ti can be used as the transducer for the thermoreflectance measurements. As will be discussed in more detail later, a metal transducer with a lower thermal conductivity is more beneficial in measuring $k_{in}$ of the substrate, especially for lowly conductive materials. The experimental system in the lab of Huazhong University of Science and Technology is shown in Fig. 1(b). This is a commercial FDTR system purchased from Fourier Scientific, designed and built by Dr. Aaron Schmidt.[32] Note that Fig. 1(b) is a simplified diagram of the system as some features designed for FDTR but irrelevant for the current method, such as a sampler for measuring the pump reference phase, are not shown here. This system uses a 488 nm blue laser as the pump beam and a 532 nm green laser as the probe beam, at which wavelength the metal Au has a very high thermoreflectance coefficient and can be used as the transducer to achieve a high SNR. Besides, this system uses a balanced photodetector, which has a large transimpedance gain and can easily achieve low noise and significantly reduce the signal-acquisition time for the same level of SNR compared to the first system. The typical laser spot size $w_0$ (the root-mean-square average of the $1/e^2$ radii of the pump and the probe laser spots) of the first and the second system is ~12 μm and ~4 μm, respectively.

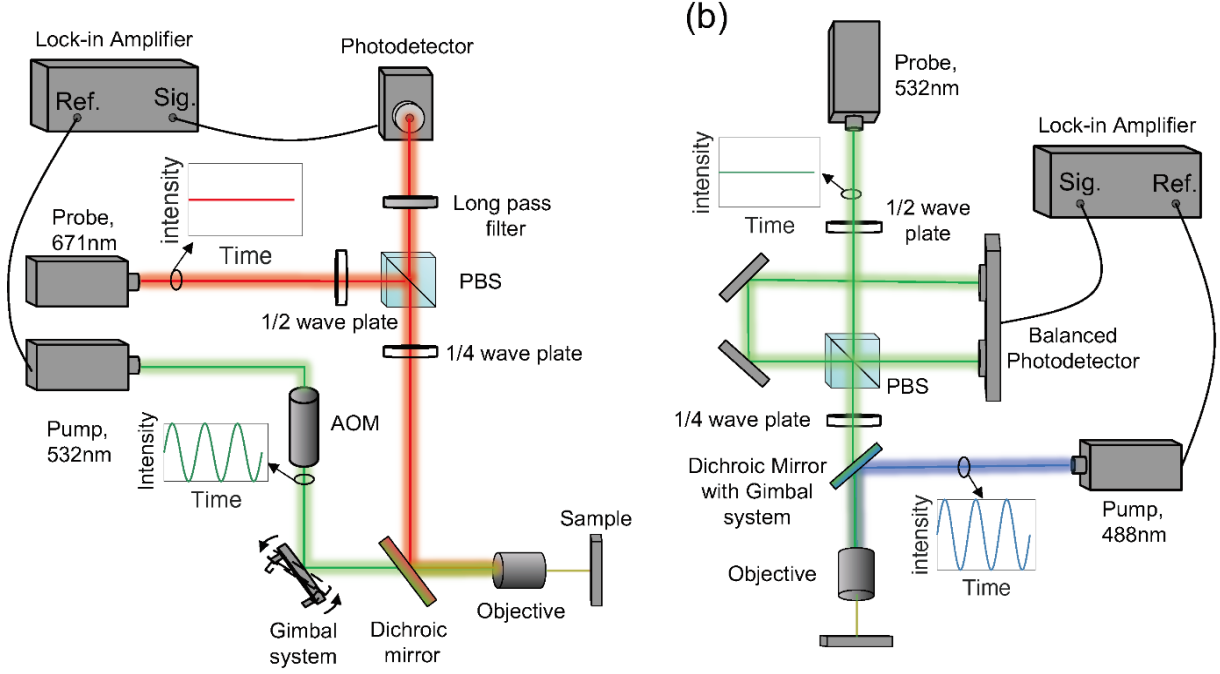

**Fig. 1.** Schematic of two experimental setups for the improved spatial-scan thermoreflectance experiments. (a) The system in the lab of the University of Pittsburgh. (b) The system in the lab of Huazhong University of Science and Technology.

In the new SSTR experiments, both the phase signal $\varphi$ and the amplitude $A$ are recorded as a function of the offset distance $x_c$ between the pump and probe spots measured using a low modulation frequency $f$. The frequency $f$ is recommended to be low enough to induce a long in-plane thermal diffusion length $d_{f,x}$ at least 3 times the laser spot size $w_x$ along the scanning direction, $d_{f,x} \geq 3w_x$. Here $d_{f,x}$ is defined as $d_{f,x} = \sqrt{k_{xx}/\pi f C}$, with $k_{xx}$ being the directional in-plane thermal conductivity of the substrate in the scanning direction $x$ and $C$ being the volumetric heat capacity of the substrate. This criterion usually guarantees a decent SNR for the phase signal even at a large offset distance of $x_c = 4w_x$ so that a phase gradient $d\varphi/dx_c$, which is essential for measuring $k_{xx}/C$ of the sample, can be reliably extracted from the phase signal in the offset range $2w_x < x_c < 4w_x$. In practice, the phase difference $\Delta\varphi = \varphi - \varphi(x_c = 0)$ as a function of $x_c$ is used to derive the in-plane thermal conductivity $k_{xx}$ in the scanning direction with $C$ as a known input, and the normalized amplitude signal $A_{norm} = A/A(x_c = 0)$ as a function of $x_c$ is used to derive the laser spot size $w_x$ by best-fitting the measured signals with a thermal model prediction. The thermal model of anisotropic heat conduction in multilayered systems has been well developed for beam-offset FDTR in the literature[22, 24] and can be directly adopted here. An example of the processed $\Delta\varphi$ and $A_{norm}$ signals for a 100 nm Ti/silica sample measured using a modulation frequency of 150 Hz and a laser spot size of 11.5 μm is shown in Fig. 2. By fitting the $\Delta\varphi$ signals in Fig. 2 (a) and with the volumetric heat capacity taken as $C = 1.66 \times 10^6$ J/(m$^3 \cdot$ K) for silica from the literature[33], the in-plane thermal conductivity of silica along the scanning direction was determined as 1.4 W/(m · K), in excellent agreement with the literature[34]. Meanwhile, the laser spot size was determined as 11.5 μm by fitting the $A_{norm}$ signals in Fig. 2 (b).

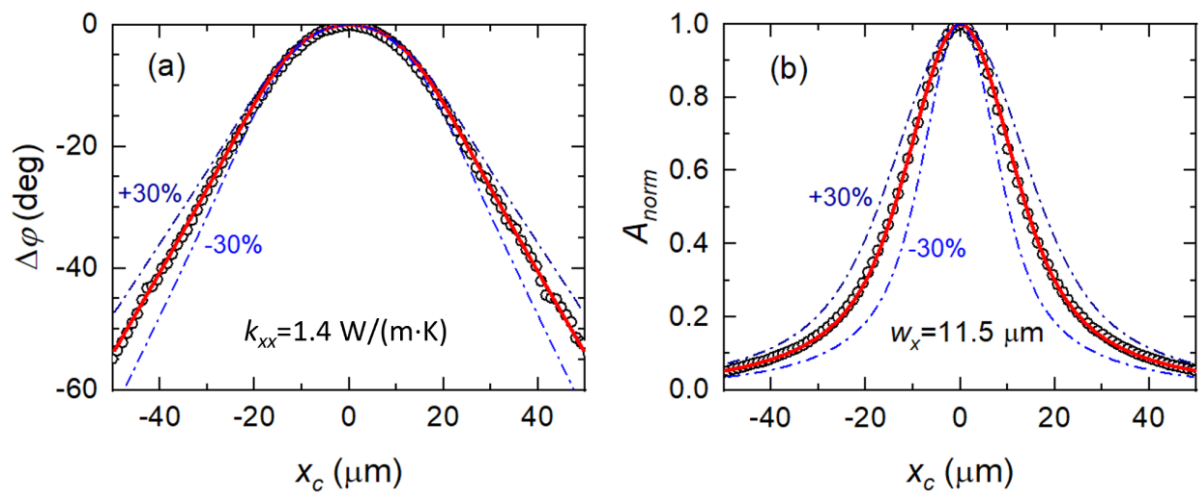

**Fig. 2.** Experimental data and model fitting for a 100 nm Ti/silica sample measured using a modulation frequency of 150 Hz and a laser spot size of 11.5μm. (a) The phase difference signal $\Delta\varphi$, from which the in-plane thermal conductivity of the silica substrate was fitted as 1.4 W/(m · K). The dash-dot curves are the ±30% bounds of the best fitted $k_{xx}$. (b) The normalized amplitude $A_{norm}$, from which the laser spot size long the offset direction was fitted as 11.5 μm. The dash-dot curves are the ±30% bounds of the best fitted $w_x$.

One novelty of the current work is the newly proposed approach of data processing. By analyzing the phase difference signal $\Delta\varphi$ rather than the absolute phase signal $\varphi$, this new SSTR method exempts the necessity of characterizing the reference phase of the lock-in amplifier and thus is more convenient to practice. Another important feature of this method is that the laser spot size can be determined at the same time as the thermal conductivity measurement rather than requiring a separate experiment to be pre-determined, which not only simplifies the experiment but also increases the reliability since the effect of any chance that the laser spot size should change between different experiments due to unintentional laser beam defocusing could be eliminated. Most importantly, the breakthrough of this method is that it successfully measures $k_{in}$ as low as $1.4 \text{ W/(m} \cdot \text{K)}$ with great accuracy, breaking the current limit of $k_{in} > 10 \text{ W/(m} \cdot \text{K)}$ measurable by TDTR and FDTR.

Although the whole range of the $\Delta\varphi$ signal was fitted to derive $k_{xx}$ in Fig. 2(a), we essentially rely on the phase gradient $\nabla\varphi = \text{d}\Delta\varphi/\text{d}x_c$ in the long offset range $x_c > 2w_x$ to determine $k_{xx}$, as demonstrated by the $\pm 30\%$ bounds of the best-fitted $k_{xx}$ value in Fig 2(a). To quantify the sensitivity of the phase signal to parameter $\alpha$, we define the sensitivity coefficient of the phase as

$$S_\alpha^{\nabla\varphi} = \frac{\alpha}{\nabla\varphi}\frac{\partial \nabla\varphi}{\partial \alpha}. \qquad (1)$$

Similarly, the sensitivity of the $A_{norm}$ signal to $w_x$ essentially comes from the full width at half maximum (FWHM) of $A_{norm}$, as demonstrated by the $\pm 30\%$ bounds of the best-fitted $w_x$ value in Fig 2(b). We define the sensitivity coefficient of the amplitude signal to parameter $\alpha$ as $S_\alpha^{FWHM}$, using the same formula as in Eq. (1), only that the signal $\nabla\varphi$ is replaced by the FWHM of $A_{norm}$.

Figure 3 shows the sensitivity coefficients $S_\alpha^{\nabla\varphi}$ and $S_\alpha^{FWHM}$ of the 100 nm Ti/silica sample measured using $f = 150 \text{ Hz}$ and $w_x = 11.5 \text{ μm}$ when scanning along the $x$ direction. All the parameters of the thermal system are analyzed, which include the anisotropic thermal conductivities $k_{xx}, k_{yy}, k_{zz}$ and the volumetric heat capacity $C$ of both the metal layer and the substrate, the thickness of the metal layer $h_m$, the interface thermal conductance $G$ between the metal layer and the substrate, and the laser spot sizes $w_x$ and $w_y$ in the *x*- and *y*-direction, respectively. For simplicity, the off-diagonal terms of the thermal conductivity tensor $k_{xy}, k_{xz}, k_{yz}$ are not shown here as they are bound to be 0 for orthogonal crystals with the principal *a*, *b*, and *c*-axis directions aligned with the *x*-, *y*-, and *z*-axis of the coordinate system, respectively. Figure 3 shows that among all the parameters, the phase gradient signal $\nabla\varphi$ is dominantly sensitive to the directional in-plane thermal diffusivity $k_{xx}/C$ of the substrate in the scanning direction *x*, with a sensitivity of ~0.5. The $\nabla\varphi$ signal is also slightly sensitive to the laser spot size $w_x$ along the scanning direction, with a sensitivity of ~0.15. The sensitivity coefficients of the $\nabla\varphi$ signal to all the other parameters are below 0.05, rendering those parameters unimportant to the measurements. On the other hand, the FWHM of $A_{norm}$ is dominantly sensitive to the laser spot size $w_x$ in the scanning direction, with a sensitivity of ~0.8. The FWHM signal is also slightly sensitive to $w_y$ and $k_{xx}$ of the substrate, with sensitivity coefficients of ~0.1 to both parameters. The sensitivity coefficients of the FWHM signal to all the other parameters are below 0.05, rendering those parameters unimportant to the measurements.

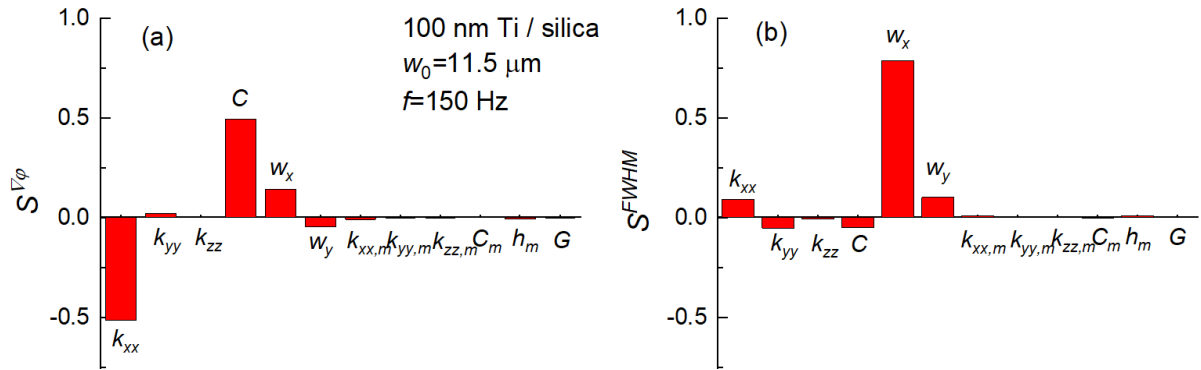

**Fig. 3.** Sensitivity coefficients of (a) the phase gradient signal $\nabla\varphi$ and (b) the FWHM of $A_{norm}$ to different parameters of a 100 nm Ti/silica sample measured by the improved spatial-scan thermoreflectance method using a modulation frequency of 150 Hz and a laser spot size of 11.5 µm. In this analysis, the scanning direction is along $x$. The analyzed parameters include the anisotropic thermal conductivities $k_{xx}, k_{yy}, k_{zz}$ and the volumetric heat capacity $C$ of both the metal layer and the substrate, the thickness of the metal layer $h_m$, the interface thermal conductance $G$ between the metal layer and the substrate, and the laser spot sizes $w_x$ and $w_y$ in the $x$- and $y$-direction, respectively. The parameters with a subscript "*m*" are for the metal transducer layer, while the unspecified parameters are for the substrate.

An iterative fitting process is used to extract $k_{xx}$ and $w_x$ simultaneously from the measured signals. (Although the phase signals are essentially sensitive to $k_{xx}/C$, we focus on measuring $k_{xx}$ with $C$ as input since in most cases the thermal conductivity is of prime interest while the heat capacity value can be found from the reference database.) As the initial step, assumption of in-plane isotropy with $k_{xx} = k_{yy} = k_{zz}$ and $k_{xy} = k_{xz} = k_{yz} = 0$ is made since the $\Delta\varphi$ signals are dominantly sensitive to $k_{xx}$. We also assume the laser spot to be circular with $w_y = w_x$ since the sensitivity of the FWHM signal to $w_y$ is much smaller than that to $w_x$. We have thus reduced eight unknown parameters $(k_{xx}, k_{yy}, k_{zz}, k_{xy}, k_{yz}, k_{xz}, w_x, w_y)$ to only two ($k_{xx}$ and $w_x$). An iterative fitting approach is necessary to extract both $k_{xx}$ and $w_x$ by fitting $\Delta\varphi$ and $A_{norm}$ signals simultaneously, i.e., the $\Delta\varphi$ signals are first fitted for $k_{xx}$ with an assumed $w_x$ value as input; the $A_{norm}$ signals are then fitted for $w_x$ with the previously fitted $k_{xx}$ value as input; the $\Delta\varphi$ signals are then fitted again with the latest $w_x$ value to have the $k_{xx}$ value updated, and so on until both the $\Delta\varphi$ and $A_{norm}$ signals could be fitted well using the same set of $k_{xx}$ and $w_x$ values.

This fitting process is sufficient for transversely isotropic materials. When measuring the full in-plane thermal conductivity tensor of transversely anisotropic materials, the above-described fitting process needs to be iteratively repeated to simultaneously fit three sets of signals measured in different scanning directions, details of which will be demonstrated through a simulated numerical experiment in Section 3.2.1.

*2.2. Uncertainty analysis*

The uncertainties of the fitted $k_{xx}$ and $w_x$ values could be estimated using a full error propagation formula. This formula fully accounts for the uncertainties of all the input parameters as well as the quality of the least-squares fitting of the experimental data and the experimental noise. Similar applications of this full error propagation formula were also found in the previous TDTR and FDTR experiments.[25, 35-38]

In the present case, since $k_{xx}$ is extracted from the $\Delta\varphi$ signal and $w_x$ is extracted from the $A_{norm}$ signal through an iterative fitting process, this fitting process can be mathematically expressed as seeking to minimize the squared difference between the experimental data and the model predictions:

$$\psi(\boldsymbol{U}) = \sum_{i=1}^{N}\left(y(x_{c,i}) - g(\boldsymbol{U},\boldsymbol{P},x_{c,i})\right)^2, \quad (3)$$

where $y(x_{c,i})$ is the experimental signal, either $\Delta\varphi$ or $A_{norm}$, at the $i$th offset spot $x_{c,i}$, and $g$ is the corresponding value evaluated by the thermal model; $\boldsymbol{U}$ and $\boldsymbol{P}$ are the column vectors of the unknown properties and the control parameters, respectively. For the current case, we have $\boldsymbol{U} = (k_{xx})^T$ and $\boldsymbol{P} = (k_{xx,m}, k_{yy,m}, k_{zz,m}, C_m, h_m, G, k_{yy}, k_{zz}, C, w_x, w_y)^T$ when fitting the $\Delta\varphi$ signals and $\boldsymbol{U} = (w_x)^T$ and $\boldsymbol{P} = (k_{xx,m}, k_{yy,m}, k_{zz,m}, C_m, h_m, G, k_{xx}, k_{yy}, k_{zz}, C, w_y)^T$ when fitting the $A_{norm}$ signals. The uncertainties of $k_{xx}$ and $w_x$ are also estimated using an iterative process similar to the procedure of extracting $k_{xx}$ and $w_x$.

Since the input parameters have uncertainties, each input parameter has a normal distribution with the mean value being its nominal value and two times the standard deviation $2\sigma$ being its uncertainty. Let us denote $\boldsymbol{P}^*$ as a random group of the possible control parameters and $\boldsymbol{U}^*$ as the corresponding group of fitting parameters that make the best fit. The uncertainties of the unknown properties could be revealed from the distribution of all the possible $\boldsymbol{U}^*$. The covariance matrix of $\boldsymbol{U}^*$ could be explicitly expressed after some analytical derivation, details of which can be found in Ref. [35]. The covariance matrix $\text{Var}[\boldsymbol{U}^*]$ takes the format of

$$\text{Var}[\boldsymbol{U}^*] = \begin{pmatrix} \sigma_{u_1}^2 & \text{cov}[u_1, u_2] & \dots \\ \text{cov}[u_2, u_1] & \sigma_{u_2}^2 & \dots \\ \vdots & \vdots & \ddots \end{pmatrix}, \tag{4}$$

where the elements on the principal diagonal are the variance of the unknown parameters. The standard deviation of each unknown parameter can be simply determined as the square root of its variance, and the uncertainty of the unknown parameter is two times its standard deviation.

In practice, we only analyze the experimental data away from the pump spot center with $|x_c| > w_x$ because the data around the center are bound to be near 0 for $\Delta\varphi$ and 1 for $A_{norm}$ and have little sensitivity to the unknown properties. Assuming an uncertainty of 10% for $k_{xx,m}$, $k_{yy,m}$, $k_{zz,m}$, $k_{yy}$, and $k_{zz}$, respectively, 20% for $G$, 5% for $h_m$, and 3% for $C_m$, $C$, and $w_y$, respectively, using the full error propagation formula we calculate that these errors of the control parameters would propagate to an uncertainty of 3% for $k_{xx}$ and 0.5% for $w_x$ for the silica data in Fig. 2. In addition to that, the experimental noise for the silica data in Fig. 2 would contribute another 1.9% uncertainty for $k_{xx}$ and 1.4% for $w_x$. The total uncertainty for $k_{xx}$ of silica is thus $\sqrt{3^2 + 1.9^2}\% = 3.5\%$ ($k_{xx} = 1.4 \pm 0.05$ W/(m·K)) and the uncertainty of $w_x$ is 1.5% ($w_x = 11.5 \pm 0.2$ μm).

Note that the well-developed TDTR and FDTR techniques typically have an uncertainty of ~10% in determining the through-plane thermal conductivity and an even larger uncertainty in determining the in-plane thermal conductivity. Due to the MHz-range high modulation frequencies used in TDTR and FDTR, the phase signals are also highly sensitive to the heat capacitance $h_m C_m$ of the metal transducer and the metal-substrate interface thermal conductance $G$, the uncertainties of which would propagate and cause a large uncertainty in determining the thermal conductivity of the substrate. This problem, however, is absent in this new SSTR method. Because a low enough modulation frequency ensuring $d_{p,x} \geq 3w_x$ is used for the measurement, the phase gradient and the normalized amplitude signals acquired in the SSTR measurements are not sensitive to the transducer layer or the metal-substrate interface (see Fig. 2 for an example), thus significantly improving the uncertainty of the measured thermal conductivity.

*2.3. Some other factors to consider*

*2.3.1. Effect of the metal transducer layer*

If the transducer layer has a much higher thermal conductivity than the substrate, heat tends to spread out in the transducer layer first before diffusing into the substrate. In this case, sensitivity to the cross-plane thermal effusivity $\sqrt{k_{zz}C}$ of the substrate and the thermal conductance $k_m h_m$ of the transducer layer increases, whereas sensitivity to the in-plane thermal diffusivity $k_{xx}/C$ of the substrate decreases. To quantify the effect of the metal transducer layer, we calculate the sensitivity of the phase gradient signal $\nabla\varphi$ as a function of metal-to-substrate thermal conductivity ratio $k_m/k_{xx}$ for two different substrates of silica and Si, each coated with a 100 nm thick artificial metal transducer layer with a varying thermal conductivity $k_m$. The heat capacity of the artificial metal layer was fixed at $C_m = 2 \times 10^6$ J/(m$^3 \cdot$K) and the laser spot size was fixed at $w_x = 10$ µm. The modulation frequency was chosen as 200 Hz for silica and 20 kHz for Si. The results are plotted in Fig. 4. From both plots, we see that the sensitivity of the $\nabla\varphi$ signal to $k_x$ of the substrate starts to reduce dramatically when $k_m/k_{xx} > 10$. Therefore, ideally, a metal transducer with a thermal conductivity $k_m < 10 k_{xx}$ is recommended to measure $k_{xx}$ of the substrate to achieve high accuracy.

Note that this conclusion is reached on the condition of a 100 nm thick metal transducer layer, which is usually required to prevent transmission of the laser beam through the metal layer and thus guarantee a surface heat flux boundary condition. Since mainly $k_m h_m$ governs the in-plane heat spreading in the transducer layer, the requirement on the highest possible thermal conductivity of the metal transducer layer can be less stringent if the metal transducer layer could be thinner, such as in the time-resolved magneto-optic Kerr effect (TR-MOKE) setups.[39, 40]

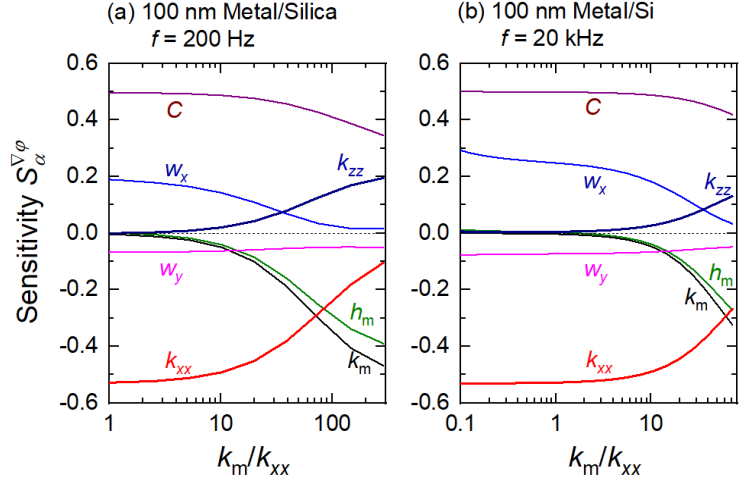

**Fig. 4.** Sensitivity coefficients of the phase gradient signal $\nabla\varphi$ to different parameters of (a) a 100 nm metal/silica system at 200 Hz, and (b) a 100 nm metal/Si system at 20 kHz, when measured using a spot size of 10 μm by the improved spatial-scan thermoreflectance method for different metal-to-substrate thermal conductivity ratios $k_m/k_{xx}$.

*2.3.2. Requirement on the sample geometry*

In the new SSTR experiments, the profile of the in-phase signal $V_\text{in}$ is mainly determined by the laser spot size whereas the profile of the out-of-phase signal $V_\text{out}$ is mainly determined by the in-plane thermal diffusion length $d_{f,x}$. We find that the $V_\text{out}$ signal would drop to <1% of its peak value at an offset distance of $x_c \geq 3d_{f,x}$. Therefore, the distance from the center of the pump spot to the sample edge should be at least $L > 3d_{f,x}$ to minimize the sample edge effect. Combining the basic criterion of $d_{f,x} \geq 3w_x$ recommended for this SSTR experiments, we require that the diameter of the sample should be $2L > 6d_{f,x} \geq 18w_x$. Therefore, if using a laser spot size of $w_0 < 5$ μm for the measurement, this improved spatial-scan thermoreflectance method can easily measure small-scale samples with a lateral diameter as small as 0.1 mm.

This method allows for measurements of both bulk and thin-film samples due to the sophisticated 3D anisotropic thermal modeling. Sensitivity analysis shows that if the film thickness $h$ is 3 times larger than $d_{f,z}$, the sensitivity of the phase signal $\Delta\varphi$ to the film thickness $h$ or the thermal properties of the substrate beneath the film would be <0.01, i.e., a 10% change in $h$ would cause <0.1% change in the phase signal. In such a case, the film sample could be considered as thermally bulk with its thickness being unimportant; otherwise, the thickness of the sample needs to be pre-determined carefully as an input parameter in the thermal model. A summary of the sample geometry requirement for this new SSTR experiment is illustrated in Fig. 5.

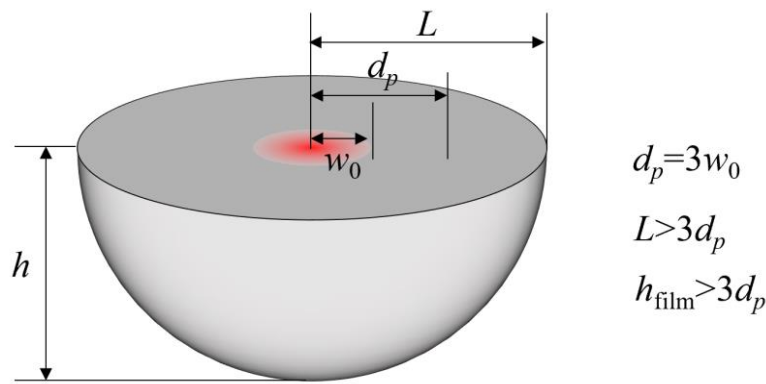

**Fig. 5.** Sample geometry required for the improved spatial-scan thermoreflectance experiment.

*2.3.3. Effect of heat loss*

One might wonder if the heat loss would affect the measurements in the SSTR experiments, especially when very low modulation frequencies are used for the measurements. The effect of heat loss on the reduction of the temperature rise can be quantified by the Biot number, which is defined as $Bi = UL_c/\sqrt{k_z k_r}$;[41] here $U$ [W/(m² · K)] is the heat loss coefficient, and $L_c$ is the characteristic length reflecting the size of the heated region. Roughly, this characteristic length can be taken as the laser spot diameter of $2w_0$.[42] The general rule is that with $Bi \ll 1$, the temperature rise in the solid is little affected by the heat loss; with $Bi \geq 0.1$, the heat loss effect on the temperature rise cannot be ignored. Let us do a simple estimation of the limiting case: given a large characteristic length of $L_c = 100$ μm in the SSTR experiments and the lowest possible thermal conductivity of $k = 0.1$ W/(m · K) for the sample, a Biot number of $Bi \geq 0.1$ would require the heat loss coefficient to be $U \geq 100 \text{ W}/(\text{m}^2 \cdot \text{K})$. Since the heat transfer coefficient of free air convection is only ~5 W/(m² · K), the heat loss effect can be safely ignored even in the atmospheric environment at room temperature.

## 3. Results and discussion

### 3.1. Validation using transversely isotropic materials: sapphire, Si, and HOPG

Figure 2 already presents the room-temperature measurement of fused silica using this new SSTR method, which yields $k_{in} = 1.4 \pm 0.05 \text{ W}/(\text{m} \cdot \text{K})$ for fused silica. This method is further validated by measuring other transversely isotropic reference samples over a broad range of $k_{in}$ values, i.e., sapphire, Si, and HOPG. Figure 6 (a, b) and (c, d) show the signals of a 100 nm Au/Si sample measured at room temperature for two different scanning directions parallel to the $x$-axis and the $y$-axis, respectively, using the same modulation frequency of 100 kHz. Independent fitting of these two sets of signals yields almost identical $k_{in}$ values for Si, i.e., 144~145 W/(m · K), although the laser spot shape is slightly elliptical with $w_x = 4.5$ μm and $w_y = 4.2$ μm and nevertheless a circular laser spot shape with $w_x = w_y$ was assumed for each fitting. Here the heat capacity of Si is taken as $C = 1.66 \times 10^6 \text{ J}/(\text{m}^3 \cdot \text{K})$;[43] therefore, this new SSTR method essentially measures an in-plane thermal diffusivity of $\alpha = 8.73 \times 10^{-5} \text{ m}^2/\text{s}$ for Si.

Figure 7 (a) shows a summary of the measured $k_{in}$ using the current method for fused silica, sapphire, Si, and HOPG at room temperature, covering a wide range from 1 to 2000 W/(m · K), and compared to the literature values. The thermal conductivities of quartz along the *a*-axis and *c*-axis, the measurement of which will be presented in detail in Section 3.2.2, are also included in this summary. All the data are also tabulated in Table I, along with the known heat capacities obtained from the literature. Overall, this new SSTR method performs remarkably well, with the measured $k_{in}$ comparing very well with the literature values.

**Table. 1.** The in-plane thermal conductivity ($k_{in}$) of some reference samples measured by the current method and compared with the literature values.

| Sample | $C$ (J/(cm$^3$ · K)) | $k$ (W/(m · K)) (Literature) | $k_{in}$ (W/(m · K)) (Current) |
|---|---|---|---|
| Fused silica | 1.66 ± 0.05[33] | 1.38 ± 0.02[34] | 1.4 ± 0.05 |
| Quartz (a-axis) | 1.94 ± 0.06[44] | 11.1 ± 0.8[45] | 11.4 ± 0.4 |
| Quartz (c-axis) | 1.94 ± 0.06[44] | 6.3 ± 0.6[45] | 6.8 ± 0.3 |
| Sapphire | 3.08 ± 0.09[46] | 35 ± 1.8[47] | 38 ± 1.5 |
| Silicon | 1.66 ± 0.05[43] | 142 ± 4[48] | 145 ± 7 |
| HOPG | 1.62 ± 0.05[49] | 1850 ± 150[50] | 1750 ± 100 |

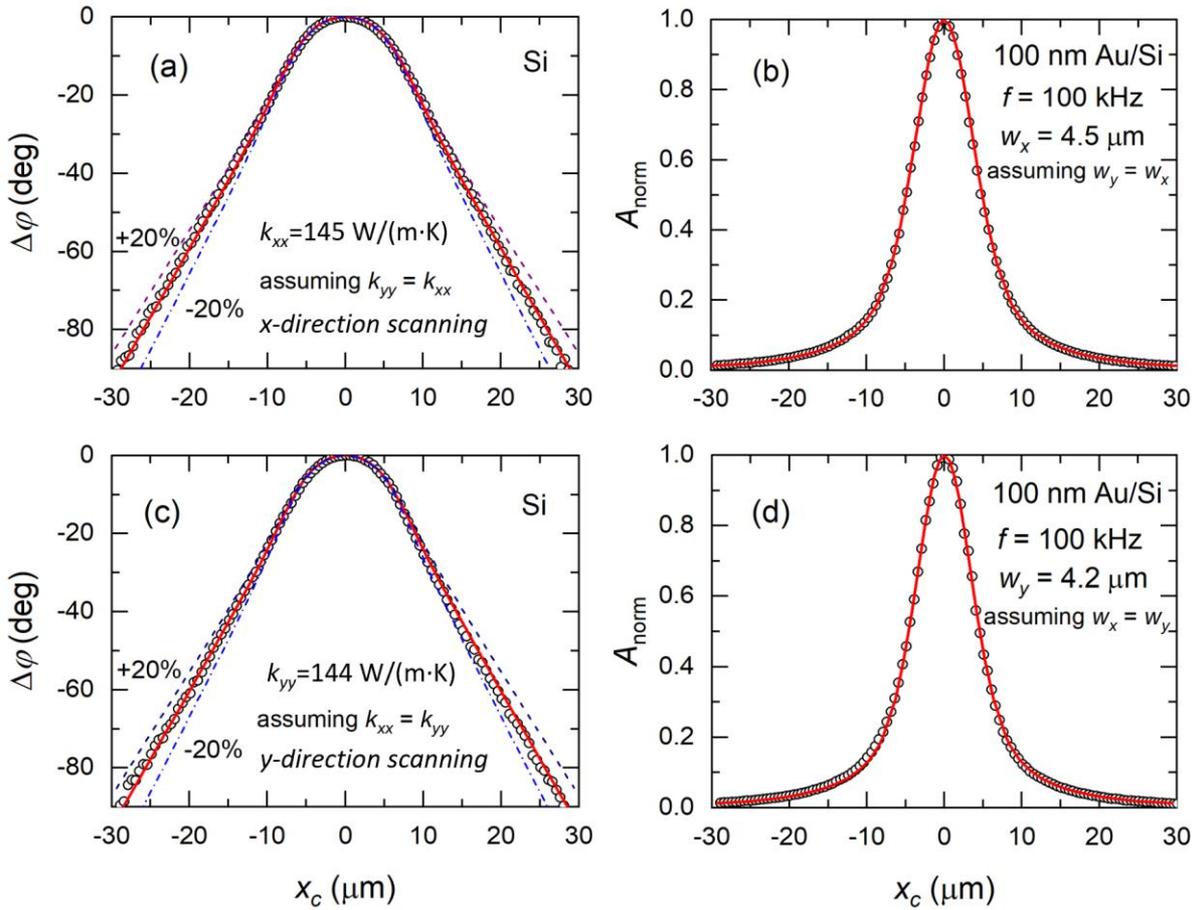

**Fig. 6.** Experimental data and model fitting for the 100 nm Au/Si sample with (a, c) phase difference $\Delta\varphi$ and (b, d) normalized amplitude $A_{\text{norm}}$ signals. The measurements are conducted using a modulation frequency of 100 kHz at room temperature. Subplots (a, b) are measured along the $x$-direction, and subplots (c, d) are measured along the $y$-direction. The laser spot sizes are individually fitted assuming $w_x = w_y$ for each fitting process.

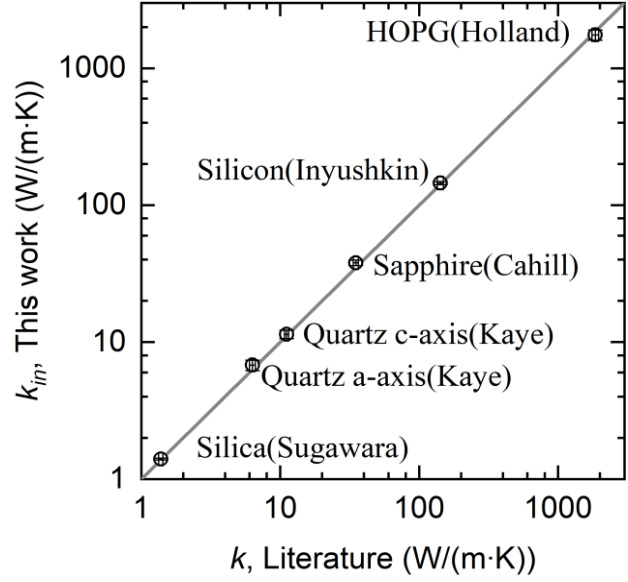

**Fig. 7.** Room-temperature measurements of $k_{in}$ of some reference samples over a broad range of $k_{in}$ values using the new spatial-scan thermoreflectance method and compared to the literature values.

*3.2. Measurements of the in-plane thermal conductivity tensor of transversely anisotropic materials*

*3.2.1. A simulated numerical experiment*

A numerical experiment is conducted to illustrate the iterative fitting process and demonstrate the unique capability of this new SSTR method for extracting both the full in-plane thermal conductivity tensor and the laser spot size, even if the laser spot shape is not ideally circular. Here we consider a hypothetical sample with a thermal conductivity tensor of

$$\mathbf{k} = \begin{bmatrix} 9 & 3 & 0 \\ 3 & 8 & 0 \\ 0 & 0 & 8 \end{bmatrix} \text{ W/(m·K)}$$ and a heat capacity of $C = 1.62 \times 10^6$ J/(m³·K), coated by a 100-nm-thick Ti layer. The interface thermal conductance, although being unimportant for the current experiment, is set as $G = 100$ MW/(m²·K). The laser spot is intentionally set as elliptical with a long radius of $w_x = 5$ µm and a short radius of $w_y = 3$ µm to account for the nonideal laser spot shapes in real experiments. To simplify the thermal modeling, the coordinate system is chosen in such a way that the long radius of the elliptical laser spot is aligned parallel to the x-axis, thus eliminating a parameter of the tilted angle of the elliptical

laser spot. The 2D in-plane thermal conductivity and laser spot size distributions are shown as curves in Fig. 8(a) and Fig. 8(b), respectively, with $k_{in}(\theta) = k_{xx}\cos^2\theta + k_{yy}\sin^2\theta + k_{xy}\sin 2\theta$ and $w(\theta) = \left(\frac{\cos^2\theta}{w_x^2} + \frac{\sin^2\theta}{w_y^2}\right)^{-1/2}$. The modulation frequency is chosen as 5 kHz, which guarantees $d_f(\theta) \geq 3w(\theta)$ for all the in-plane directions. The $\Delta\varphi$ and $A_{norm}$ signals of this system predicted by the thermal model for three different scanning directions of $\theta = 0°, 45°,$ and $90°$, presented as symbols in Fig. 8(c-h), are treated as experimental signals to be fitted.

To extract the full in-plane thermal conductivity tensor and the elliptical laser spot size, the first step is to extract $k_{xx}$ and $w_x$ by fitting the signals measured in the $\theta = 0°$ direction with the assumptions of $k_{xx} = k_{yy} = k_{zz}$, $k_{xy} = k_{xz} = k_{yz} = 0$, and $w_x = w_y$. This fitting yields $k_{xx} = 8.3$ W/(m·K) and $w_x = 4.4$ µm. With the latest value of $w_x$ as input, $k_{yy}$ and $w_y$ are then extracted by fitting the signals measured in the $\theta = 90°$ direction with the assumptions of $k_{xx} = k_{yy} = k_{zz}$, and $k_{xy} = k_{xz} = k_{yz} = 0$. This fitting yields $k_{yy} = 7.2$ W/(m·K) and $w_y = 3$ µm. Note that assuming $k_{xx} = k_{yy}$ here is more helpful for fast convergence than taking the last $k_{xx}$ value as input. We then take the latest $k_{xx}$, $k_{yy}$, $w_x$, and $w_y$ values as inputs and fit the signals measured in the $\theta = 45°$ direction for $k_{xy}$, which yields $k_{xy} = 3.9$ W/(m·K). This finishes the first round of the iteration.

For the second round of the iteration, the signals measured in the $\theta = 0°$ direction are first fitted again to have $k_{xx}$ and $w_x$ updated with the most recent values of the other parameters as inputs, which yields $k_{xx} = 9.4$ W/(m·K) and $w_x = 5$ µm. We then update $k_{yy}$ and $w_y$ by fitting the signals measured in the $\theta = 90°$ direction again with the most recent values of the other parameters as inputs, which yields $k_{yy} = 8.5$ W/(m·K) and $w_y = 3$ µm. The third step is to update $k_{xy}$ by fitting the signals measured in the $\theta = 45°$ direction again, which yields $k_{xy} = 3.9$ W/(m·K).

The same fitting process as the second round of the iteration can be repeated as many times as necessary until the fitted $k_{xx}$, $k_{yy}$, $k_{xy}$, $w_x$, and $w_y$ values converge. Figure 8(i) shows the fitted $k_{xx}$, $k_{yy}$, and $k_{xy}$ values and Fig. 8(j) shows the fitted $w_x$ and $w_y$ values as a function of the number of iterations, with the horizontal lines being the target values. By the fourth round of iteration, the thermal conductivity tensor has converged to values differing by <3% from the target values. The laser spot sizes have converged to the target values by the second round of iteration. The extracted $k_{in}$ tensors and the fitted laser spot sizes after the first and the third round of iteration are shown as dashed and dash-dot curves in Figure 8(a) and 8(b), respectively, as a comparison with the target values. The whole iterative fitting process is also illustrated in Table 2.

**Table 2.** An illustration of the step-by-step fitting process for the new spatial-scan thermoreflectance method to simultaneously extract the anisotropic in-plane thermal conductivity tensor and the elliptical laser spot sizes. The framed parameters are freely adjustable and the others are the fixed inputs. The subscript $n$ represents the parameters from the $n$th iteration of the fitting.

|  | 0° | 90° | 45° | 0° | 90° | 45° | … | 0° | 90° | 45° |
|---|---|---|---|---|---|---|---|---|---|---|
| $k_{xx}$ | $\boxed{k_{xx,1}}$ | $=\boxed{k_{yy,1}}$ | $k_{xx,1}$ | $\boxed{k_{xx,2}}$ | $k_{xx,2}$ | $k_{xx,2}$ |  | $\boxed{k_{xx,n}}$ | $k_{xx,n}$ | $k_{xx,n}$ |
| $k_{yy}$ | $=\boxed{k_{xx,1}}$ | $\boxed{k_{yy,1}}$ | $k_{yy,1}$ | $k_{yy,1}$ | $\boxed{k_{yy,2}}$ | $k_{yy,2}$ |  | $k_{yy,n-1}$ | $\boxed{k_{yy,n}}$ | $k_{yy,n}$ |
| $k_{xy}$ | 0 | 0 | $\boxed{k_{xy,1}}$ | $k_{xy,1}$ | $k_{xy,1}$ | $\boxed{k_{xy,2}}$ | … | $k_{xy,n-1}$ | $k_{xy,n-1}$ | $\boxed{k_{xy,n}}$ |
| $w_x$ | $\boxed{w_{x,1}}$ | $w_{x,1}$ | $w_{x,1}$ | $\boxed{w_{x,2}}$ | $w_{x,2}$ | $w_{x,2}$ |  | $\boxed{w_{x,n}}$ | $w_{x,n}$ | $w_{x,n}$ |
| $w_y$ | $=\boxed{w_{x,1}}$ | $\boxed{w_{y,1}}$ | $w_{y,1}$ | $w_{y,1}$ | $\boxed{w_{y,2}}$ | $w_{y,2}$ |  | $w_{y,n-1}$ | $\boxed{w_{y,n}}$ | $w_{y,n}$ |
|  | 1st iteration | | | 2nd iteration | | | … | $n$th iteration | | |

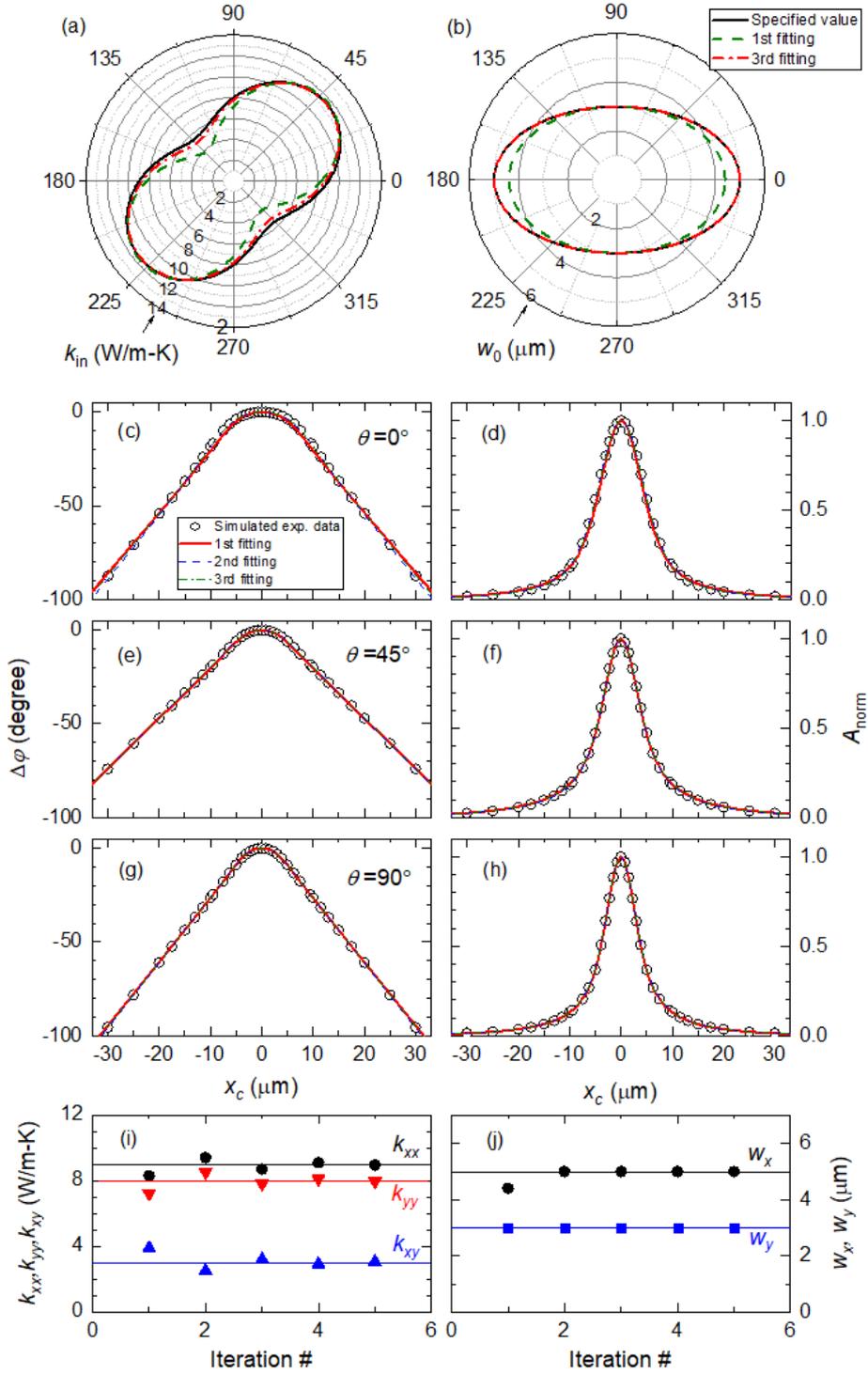

**Fig. 8.** A simulated numerical experiment to demonstrate the iterative fitting process as well as its capability in extracting anisotropic in-plane thermal conductivity tensor and elliptical laser spot size using the improved spatial-scan thermoreflectance method. (a, b) The target values of the anisotropic $k_{in}$ tensor and elliptical laser spot size (solid curves), compared to the extracted values from the first (dashed curves) and the third (dash-dot curves) round of the fitting. (c-h) The simulated experimental signals (symbols) of $\Delta\varphi$ and $A_{\text{norm}}$ for three different scanning directions of 0°, 45°, and 90°, along with their best-fitting curves. (i, j) The fitted $k_{xx}$, $k_{yy}$, $k_{xy}$, $w_x$, and $w_y$ values plotted as a function of the number of iterations. The horizontal lines are their target values, respectively.

In the above fitting process, $k_{zz}$ can be arbitrarily set as any reasonable value as its effects on the $\Delta\varphi$ and $A_{norm}$ signals are negligibly small. Note that we have aligned the *x*-axis of the coordinate system to be parallel to the principal axis of the elliptical laser spot, which reduces the parameter of the tilting angle of the elliptical laser spot and thus significantly simplifies the thermal model. Therefore, in real experiments when the laser spot shape is not ideally circular, the major and minor radii of the elliptical laser spot need to be characterized first, and then the coordinate system needs to be redefined with the *x*-axis aligned either along with or perpendicular to the principal axis of the elliptical laser spot to be compatible with the thermal model.

*3.2.2. x-cut quartz*

An x-cut quartz (<110>-oriented) sample, which is transversely anisotropic, is coated with a 100 nm Ti layer and measured by the system in Fig. 1(a) to validate the capability of this method in measuring anisotropic in-plane thermal conductivity tensor. The laser spot size is approximately 10 μm, and the modulation frequency is set as 1 kHz to meet the recommended criterion of $d_{f,x} \geq 3w_x$.

During the measurement, the sample is randomly positioned and the laser spot shape could be a tilted ellipse. Therefore, on the coordinate system defined by the sample stage, signals of 12 different scanning directions of $\theta = 0°, 30°, 60° \ldots 330°$ are first acquired and independently fitted with the assumptions of isotropic in-plane thermal conductivity and circular laser spot shape, with the preliminary results of $w(\theta)$ and $k_{in}(\theta)$ for each scanning direction plotted as the solid symbols in Fig. 9(a) and 9(b), respectively. Best-fitting these spot size data using a tilted ellipse function, the major and minor radii of the elliptical laser spot are determined as 11.6 μm and 9.8 μm, respectively, with a tilting angle of 0.23°, which is negligibly small and is thus ignored. Therefore, the coordinate system defined by the sample stage coincides with the one preferable for the thermal model.

The next step is to iteratively fit the signals acquired for three different scanning directions of $\theta = 0°, 30°,$ and $90°$ using the same fitting process described in Section 3.2.1. The fitted results have converged by the second round of the iterative fitting process thanks to the not-so-elliptical laser spot shape, with the best-fitted $k_{in}$ tensor determined as $k_{xx} = 11.3 \pm 0.4$ W/(m·K), $k_{yy} = 6.9 \pm 0.3$ W/(m·K), and $k_{xy} = 0.8 \pm 0.4$ W/(m·K), and the laser spot size as $w_x = 11.3 \pm 0.2$ μm and $w_y = 9.3 \pm 0.2$ μm. The directional laser spot profile $w(\theta) = \left(\frac{\cos^2\theta}{w_x^2} + \frac{\sin^2\theta}{w_y^2}\right)^{-1/2}$ and directional $k_{in}$ profile $k_{in}(\theta) = k_{xx}\cos^2\theta + k_{yy}\sin^2\theta + k_{xy}\sin 2\theta$ are shown as the solid curves in Fig. 9(a) and 9(b), respectively. The uncertainty of $w(\theta)$ is determined as $\eta_{w(\theta)} = \frac{1}{w(\theta)}\sqrt{(\frac{\partial w(\theta)}{\partial w_x})^2\delta_{w_x}^2 + (\frac{\partial w(\theta)}{\partial w_y})^2\delta_{w_y}^2}$, and that of $k_{in}(\theta)$ is determined as $\eta_{k_{in}(\theta)} = \frac{1}{k_{in}(\theta)}\sqrt{(\frac{\partial k_{in}(\theta)}{\partial k_{xx}})^2\delta_{k_{xx}}^2 + (\frac{\partial k_{in}(\theta)}{\partial k_{yy}})^2\delta_{k_{yy}}^2 + (\frac{\partial k_{in}(\theta)}{\partial k_{xy}})^2\delta_{k_{xy}}^2}$. They are shown by the shadowed regions in Fig. 9(a) and 9(b), respectively.

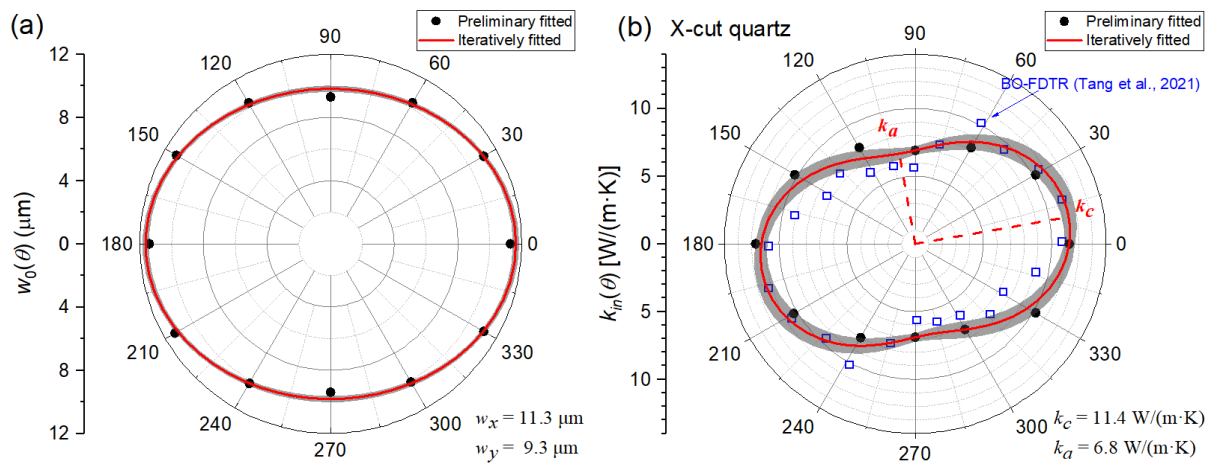

**Fig. 9.** (a) Laser spot size profile and (b) in-plane thermal conductivities of x-cut quartz from the current measurements. The solid symbols are the preliminary results fitted independently for each scanning direction assuming circular laser spot shape and isotropic $k_{in}$. The solid curves are the final results obtained through an iterative fitting of three sets of signals obtained for different scanning directions of $\theta = 0°, 30°,$ and $90°$. The directional $k_{in}$ values obtained by a recent beam-offset FDTR method[24] are also included in (b) as open symbols for comparison.

Rotating the original thermal conductivity tensor $\mathbf{k}_{in} = \begin{bmatrix} k_{xx} & k_{xy} \\ k_{yx} & k_{yy} \end{bmatrix}$ counterclockwise by an angle $\theta$ yields a new thermal conductivity tensor $\mathbf{k}'_{in}(\theta) = \begin{bmatrix} k'_{xx}(\theta) & k'_{xy}(\theta) \\ k'_{xy}(\theta) & k'_{yy}(\theta) \end{bmatrix}$, with the new off-diagonal element as $k'_{xy}(\theta) = \frac{1}{2}(k_{yy} - k_{xx}) \sin 2\theta + k_{xy}(\theta) \cos 2\theta$. Setting $k'_{xy}(\theta_0) = 0$ yields $\theta_0 = \frac{1}{2} \arctan\left(\frac{2k_{xy}}{k_{xx}-k_{yy}}\right)$. For the current measurement, $\theta_0$ is determined as $10°$. Therefore, the thermal conductivities parallel to the c-axis and the a-axis of the x-cut quartz are $k_c = k_{in}(10°) = 11.4 \pm 0.4 \text{ W/(m}\cdot\text{K)}$ and $k_a = k_{in}(100°) = 6.8 \pm 0.3 \text{ W/(m}\cdot\text{K)}$, respectively, which are in excellent agreement with the literature values[45]. As a comparison, recent measurements of the $k_{in}$ tensor of x-cut quartz by a beam-offset FDTR (BO-FDTR) method[24] are rotated by a $\theta_0$ angle and plotted as open symbols in Fig. 9(b). Overall, these two sets of measurements are consistent. However, much better quality is evident for the current measurements as they have smaller measurement uncertainty and more consistent results between different scanning directions as compared to the literature measurements by the BO-FDTR method[24].

## IV. CONCLUSIONS

In summary, a new spatial-scan thermoreflectance method has been developed to measure the in-plane thermal conductivity tensor of millimeter-scale small samples over a broad range of thermal conductivity values from 1-2000 W/(m·K), extending the current limit of the measurable $k_{in}$ by other thermoreflectance methods to as low as 1 W/(m·K). Details of this method including the experimental configuration, sensitivity analysis, and uncertainty estimation are fully described. This method has been validated by measuring several in-plane isotropic reference samples including fused silica, sapphire, silicon, and HOPG, with a typical uncertainty of ~5%. The capability of this method in measuring the in-plane thermal conductivity tensor of transversely anisotropic materials even using a slightly elliptical laser spot shape has been demonstrated through both a simulated numerical experiment and a real measurement of x-cut quartz, with the results agreeing perfectly well with the literature values.

**ACKNOWLEDGMENTS:**

P.J. acknowledges the financial support by the R.K. Mellon Postdoctoral Fellowship during his research in Pitt and the HUST startup grant #3004120159.

**DATA AVAILABILITY**

The data that support the findings of this study are available from the corresponding author upon reasonable request.